\documentclass{article}

\usepackage{arxiv}

\usepackage[utf8]{inputenc} % allow utf-8 input
\usepackage[T1]{fontenc}    % use 8-bit T1 fonts
\usepackage{url}            % simple URL typesetting
\usepackage{booktabs}       % professional-quality tables
\usepackage{amsfonts}       % blackboard math symbols
\usepackage{nicefrac}       % compact symbols for 1/2, etc.
\usepackage{microtype}      % microtypography
\usepackage{amsmath}
\usepackage{lipsum}
\usepackage{graphicx}
\graphicspath{ {./images/} }
\usepackage{tabularx,colortbl}
\usepackage[table,xcdraw]{xcolor}
\usepackage{comment}
\usepackage{booktabs}
\usepackage{amsmath}
\usepackage{stfloats}

\title{Design and Experimental Validation of UAV Swarm-Based Phased Arrays with MagSafe- and LEGO-Inspired RF Connectors}

\author{
 Bidya Debnath \\
  Mississippi State University\\
  \texttt{bd1166@msstate.edu} \\
   \And
  Mst Mostary Begum \\
  Mississippi State University\\
  \texttt{mb4198@msstate.edu} \\
  \And
 Prashant Neupant \\
  Mississippi State University\\
  \texttt{pn300@msstate.edu} \\
  \And
   Brooke E. Molen \\
  Mississippi State University\\
  \texttt{bem352@msstate.edu} \\
  \And
  Junming Diao \\
  Mississippi State University\\
  \texttt{jdiao@ece.msstate.edu} \\
}

\begin{document}
\maketitle
\begin{abstract}
This paper presents a novel UAV swarm-based phased array antenna system that leverages MagSafe- and LEGO-inspired radio frequency (RF) connectors to address key challenges in distributed phased arrays, including inter-element oscillator synchronization, localization, phase coherence, and positional accuracy. The proposed non-threaded, hands-free connectors enable precise inter-element spacing and establish a continuous, low-loss RF signal propagation path during mid-flight docking. A multi-stage optimization of the RF connector achieves a compact form factor, DC-to-RF bandwidth, and a measured insertion loss as low as 0.2\,dB. The system architecture offers scalability in gain and frequency by adjusting the array element density per UAV and UAV dimensions. Experimental results from both stationary and in-flight tests of two UAV-based phased array prototypes align closely with simulations, demonstrating robust beam steering to multiple directions. This work delivers a practical, scalable, and low-complexity platform that enables rapid deployment for next-generation airborne communications, radar, and remote sensing applications.
\end{abstract}

% keywords can be removed

\section{Introduction}
Unmanned Aerial Vehicle (UAV) swarms have emerged as a promising platform for creating airborne distributed phased arrays. In such systems, individual UAVs, each carrying a single antenna element or a multi-element subarray~\cite{yoon2017conformal,8633981}, collectively form a large-scale virtual array~\cite{namin2012analysis}. These distributed arrays minimize ground structure blockage and, owing to their flexible and dynamic nature, enable advanced functionalities such as adaptive beamforming, spatial filtering, and high-gain directional transmission~\cite{10632038}. These capabilities enhance coverage, improve interference rejection, and increase communication reliability~\cite{nanzer2021distributed}. UAV swarm-based phased array systems have found potential applications in wireless communications~\cite{zeng2016wireless,huo2019distributed}, remote sensing~\cite{9947079,klemas2015coastal,kurum2021uas}, radar~\cite{9758040,10632038,9048820}, and mobile base stations~\cite{choi2025independent,7589913}.

%Their ability to extend communication range, reduce line-of-sight (LoS) blockage~\cite{10409175,duangsuwan2021measurement}, and provide dynamic reconfigurability makes them particularly valuable for missions such as disaster response~\cite{saraereh2020performance}, defense surveillance~\cite{brust2021swarm,wang2024survey}, environmental monitoring~\cite{fascista2022toward}, and ad hoc wireless networks~\cite{brown2004ad,rohde2013ad}.

Traditional solid phased array systems, with all elements physically connected to a common transmitter source or receiver load, inherently avoid synchronization issues by sharing a coherent source and maintaining continuous RF paths through threaded connectors. In contrast, UAV swarm-based distributed arrays lack such mid-air physical interconnections, making it extremely challenging to coherently combine signals from spatially separated array elements. These challenges include frequency offsets from independent local oscillators~\cite{ouassal2021decentralized}, dynamic inter-UAV positioning, and strict phase/time synchronization requirements~\cite{diao2019experimental,nanzer2021distributed}. Moreover, aerial motion and environmental disturbances introduce spatiotemporal variations~\cite{chatterjee2017study}, necessitating precise localization~\cite{petko2009positional} and adaptive topology control~\cite{li2021design}. For instance, maintaining constructive interference requires path differences within \(\lambda/10\) (approximately \(36^\circ\) phase error), which translates to positional tolerances of \(1\,\mathrm{cm}\) at \(3\,\mathrm{GHz}\). Timing error tolerances are equally stringent at \(33.3\,\mathrm{ps}\)~\cite{8602642}. Additionally, reducing network delays, preventing eavesdropping and jamming, and avoiding UAV collisions further increase system complexity. These strict requirements constrain coherent signal addition and scalability, making the practical implementation of distributed beamforming systems particularly challenging~\cite{nanzer2021distributed}.

Several strategies have been explored to mitigate these challenges. Closed-loop beamforming relies on destination feedback to iteratively correct phase and position errors; however, its steering agility is limited by feedback latency, and the convergence time increases drastically with array size and carrier frequency~\cite{9077393,9958941}. Open-loop approaches aim to achieve phase alignment without relying on feedback from the receiver. These methods assume that each transmitter has knowledge of its own location and clock behavior and can coordinate based on shared reference information. Techniques based on Global Positioning System (GPS) and real-time kinematic (RTK) positioning~\cite{morrison2024robust} offer a simpler architecture, but remain vulnerable to receiver noise, clock drift, and multipath effects, which collectively constrain scalability at radio frequencies~\cite{9597559}. Despite these endeavors, no existing technique simultaneously achieves the sub-wavelength positional precision, low latency, and moderate system complexity required for large-scale UAV swarm arrays operating across a wide RF bandwidth.

\begin{figure}[h]
\centering
{\includegraphics[width=0.6\textwidth]{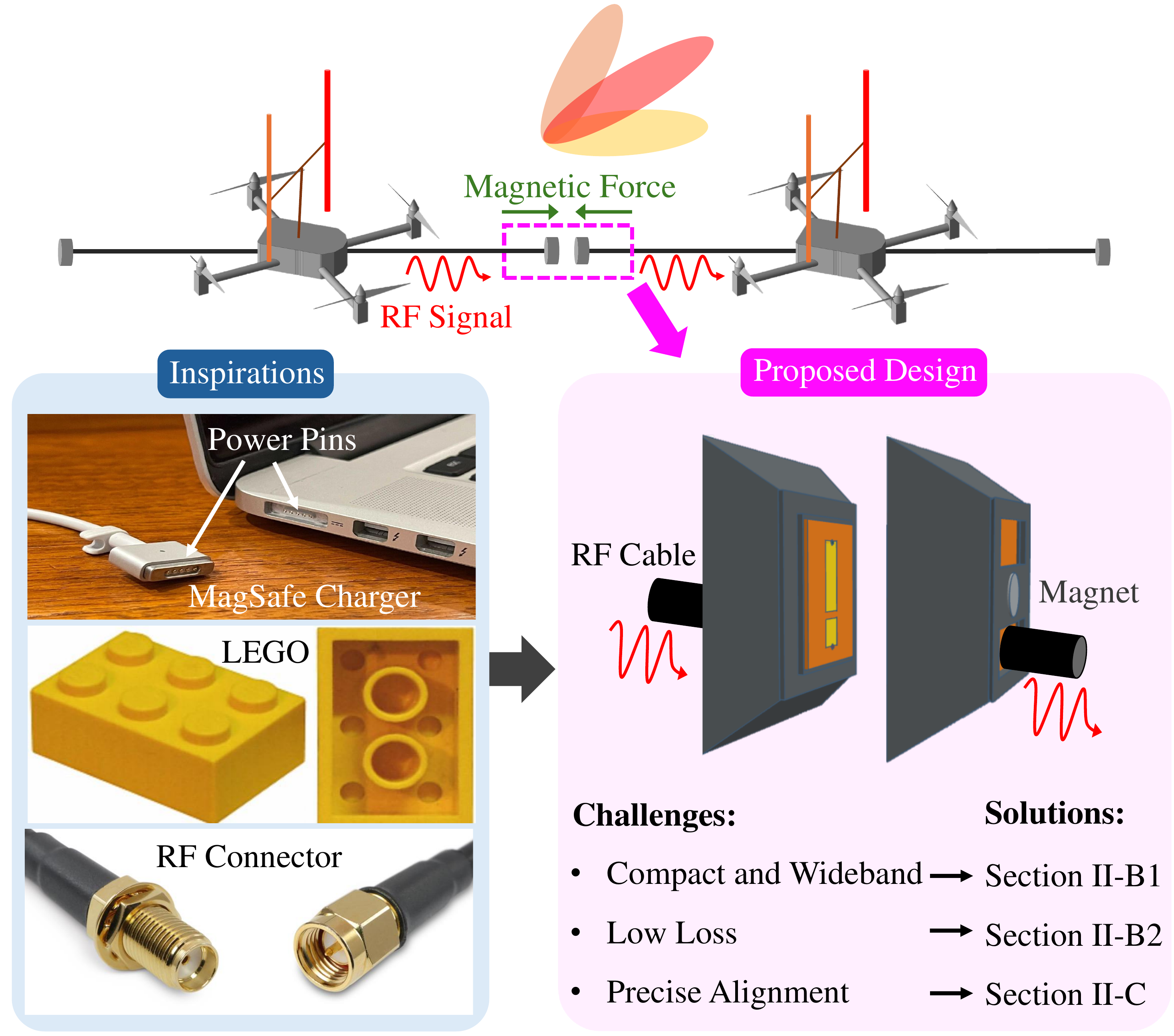}}
\caption{Design motivation and concept of the proposed RF connector.}
\label{fig:proposed_connector}
\end{figure}

In this paper, we propose a novel UAV swarm-based phased array architecture that eliminates the need for synchronization and localization by employing airborne, physically connected UAVs. The system utilizes non-threaded and hand-free RF connectors (see Fig.~\ref{fig:proposed_connector}) to enable autonomous mid-flight docking of UAVs. Once docked, the UAVs form a mechanically rigid phased array with minimal positional error. These RF connectors provide low-insertion-loss signal transmission from DC-to-RF bandwidth, supporting seamless RF coupling between UAV-mounted elements and allowing the distributed array to operate as a single coherent system. The resulting architecture is scalable and low-latency, offering improved signal integrity and deployment flexibility. This design provides a robust solution for dynamic, high-performance airborne wireless communication systems, with applications in disaster response, radar, relay networks, and robust connectivity in complex environments.

In this work, we present a phased array system optimized for UAV swarm deployment. The study encompasses the design and validation of an RF connector with magnetic interlocking and guided alignment structures, analysis of array gain as a function of UAV count, and exploration of operating frequency scalability through adjustments to connector design, UAV dimensions, and the number of antenna elements per UAV. Beamsteering capability is assessed via phase control, and radiation patterns are experimentally verified through anechoic chamber measurements. Finally, we demonstrate the feasibility of mid-flight UAV docking and undocking, validating robust beamsteering performance under dynamic flight conditions.

The remainder of this paper is organized as follows: Section 2 discusses the motivation and design evolution of the proposed RF connector. Section 3 describes the phased array design and scalability analysis. Section 4 presents the fabrication process and experimental validation using both non-flying and flying prototypes of the UAV swarm-based phased array. Finally, Section 5 concludes with a summary of key findings, advantages, and potential applications.

\section{Proposed RF Connector Design}
In this section, we present and analyze a novel RF connector design inspired by practical magnetic interface technologies. We begin by outlining the key design motivations, focusing on minimizing insertion loss, addressing bandwidth limitations, and improving alignment accuracy, along with their corresponding solutions. The design evolution is evaluated using the High-Frequency Structure Simulator (HFSS, Ansys Inc.), a full-wave finite element analysis tool. We then examine how the proposed design achieves reduced insertion loss and enhanced bandwidth. Furthermore, we assess the effectiveness of precise mechanical alignment facilitated by alignment brackets secured through magnetic attachment, and discuss their impact on the overall performance of the RF connector.

\begin{figure*}[htbp]
    \centering
    \includegraphics[width=1\textwidth]{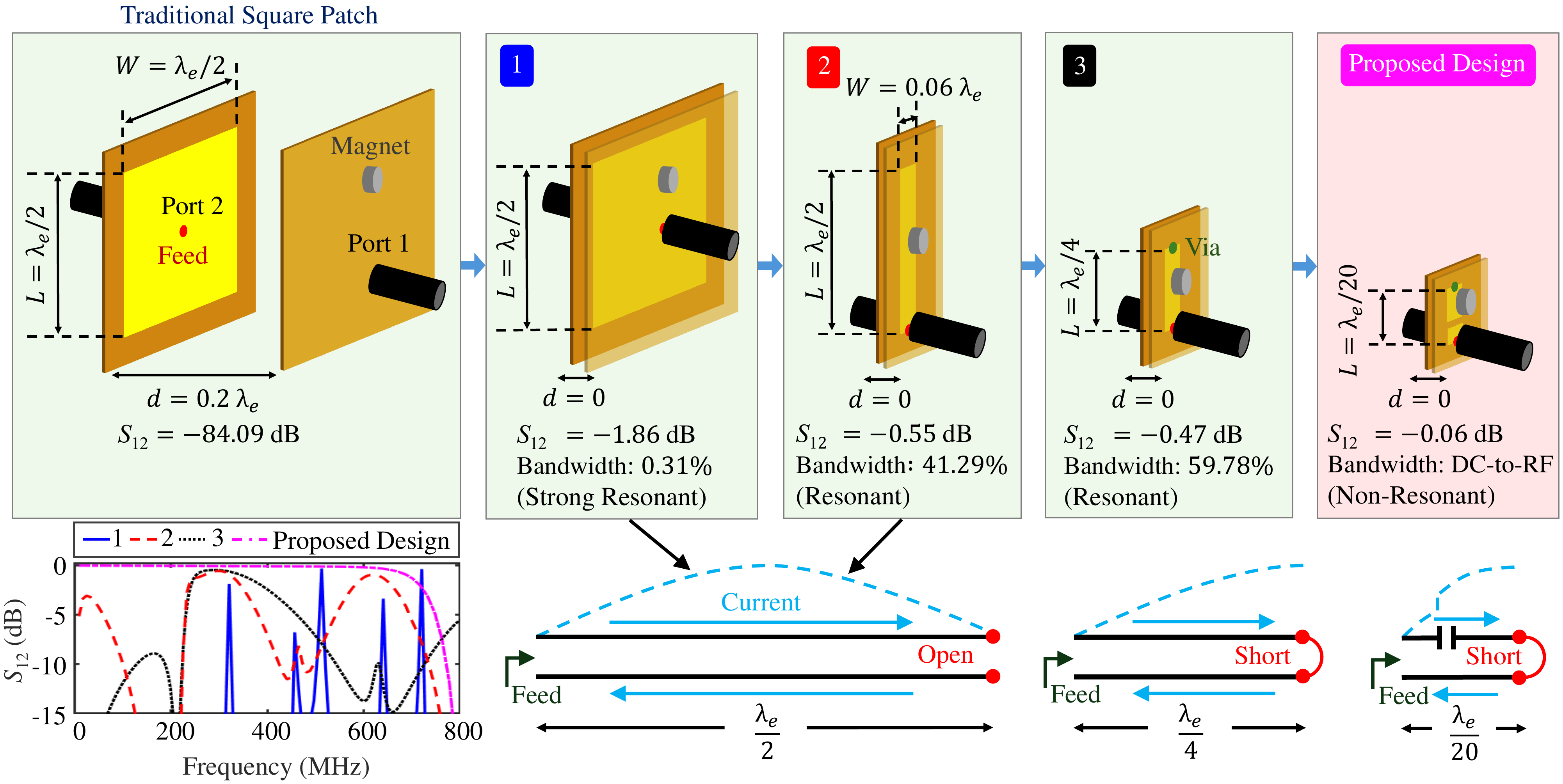}
    \caption{Design evolution of the RF connector showing the progression from the traditional square patch to the proposed compact structure with improved $S_{12}$ and wider bandwidth.}
    \label{fig:connector_design_trend}
\end{figure*}

\subsection{Motivation}
The proposed RF connector design is motivated by the challenges of localization, local oscillator synchronization, and phase coherence in UAV swarm-based distributed phased arrays. Conventional phased arrays on static platforms typically overcome these issues using physical RF interconnections, such as coaxial cables and threaded connectors. However, threaded RF connectors require a twisting motion to secure the connection, which is infeasible for establishing links mid-flight between UAV-mounted components of a phased array antenna. To address these limitations, the proposed design draws inspiration from three key sources, as illustrated in Fig.~\ref{fig:proposed_connector}. 

First, Apple's MagSafe technology demonstrates a magnetic, contact-based interface that enables effortless alignment and reliable electrical connectivity without manual intervention~\cite{ali2021efficient}. Inspired by this, the proposed RF connector integrates embedded magnets to facilitate automatic mid-flight alignment and tool-free, contact-based RF interconnection, eliminating the need for threaded coupling or manual assembly. Second, LEGO’s stud-and-socket design inspires the incorporation of a guided mechanical structure, ensuring passive, repeatable alignment, precise inter-element positioning, and structural rigidity. Finally, traditional RF connectors achieve low-loss transmission across a wide bandwidth through physical contact. The proposed design seeks to retain this wideband performance within a compact, non-threaded, and manual-intervention-free form factor, making it suitable for mid-flight docking and enabling UAVs to assemble into a coherent swarm-based phased array comparable to a ground-based system.

\subsection{MagSafe-Inspired RF Connector Design}
The design evolution of the proposed RF connector in Fig.~\ref{fig:proposed_connector} is illustrated in Fig.~\ref{fig:connector_design_trend}. We followed a systematic progression from a strong resonant traditional square patch structure toward the proposed non-resonant, compact and wideband configuration presented in this study. Each stage represents a purposeful modification aimed at achieving  $S_{12}$ values closer to $0$ dB and widening the $3$ dB bandwidth.

\subsubsection{Compact and Wideband Design}
\label{sec:Compact_Wideband}
The RF connector is designed for an operating frequency of \(300\,\mathrm{MHz}\). As shown in Fig.~\ref{fig:connector_design_trend}, the design evolution progresses through five distinct stages, starting from the traditional square patch and culminating in the final proposed design. In each stage, the width \(W\) and length \(L\) of the patch antenna in the RF connector are expressed relative to the effective wavelength \(\lambda_e\) on an FR4 substrate with a relative permittivity of 4.4 and a loss tangent of 0.02. The parameter \(d\) represents the air gap distance between the connectors.

We begin the design process with a square patch of dimensions \(W = L = \lambda_e / 2\) and an air gap \(d = 0.2\lambda_e\), which exhibits poor coupling performance, resulting in an \(S_{12}\) of approximately \(-84.09\,\mathrm{dB}\). Reducing the air gap to \(d = 0\) in stage 1 significantly improves the transmission coefficient to \(S_{12} = -1.86\,\mathrm{dB}\). However, the bandwidth remains narrow at only \(0.31\%\), a limitation attributed to the strong resonant behavior of the \(\lambda_e/2\) square patch. In stage 2, further geometric optimization involves reducing the width to \(W = 0.06\lambda_e\), which effectively weakens the resonance and increases the bandwidth to \(41.29\%\). The choice of a patch length equal to \(0.5\lambda_e\) can be explained by the analogy to an open-circuited transmission line, where resonance occurs at half-wavelength due to standing wave formation in the current distribution.

To further reduce the dimensions, stage~3 transitions from an open-circuit transmission line to a short-circuit configuration by introducing a metallic via that connects the ground plane to the patch. This modification effectively reduces the required patch length \(L\) from \(\lambda_e/2\) to \(\lambda_e/4\) and expands the \(3\,\mathrm{dB}\) bandwidth to \(59.78\%\). By providing a well-defined boundary condition, the via confines the electromagnetic fields, ensuring that the resonant frequency of stage~3 remains consistent with that of stage~2. This approach enables a more compact design and wider bandwidth compared to the earlier open-circuit configurations.

\begin{figure}[htp!]
      \begin{center}
      \includegraphics[width=0.7\textwidth]
{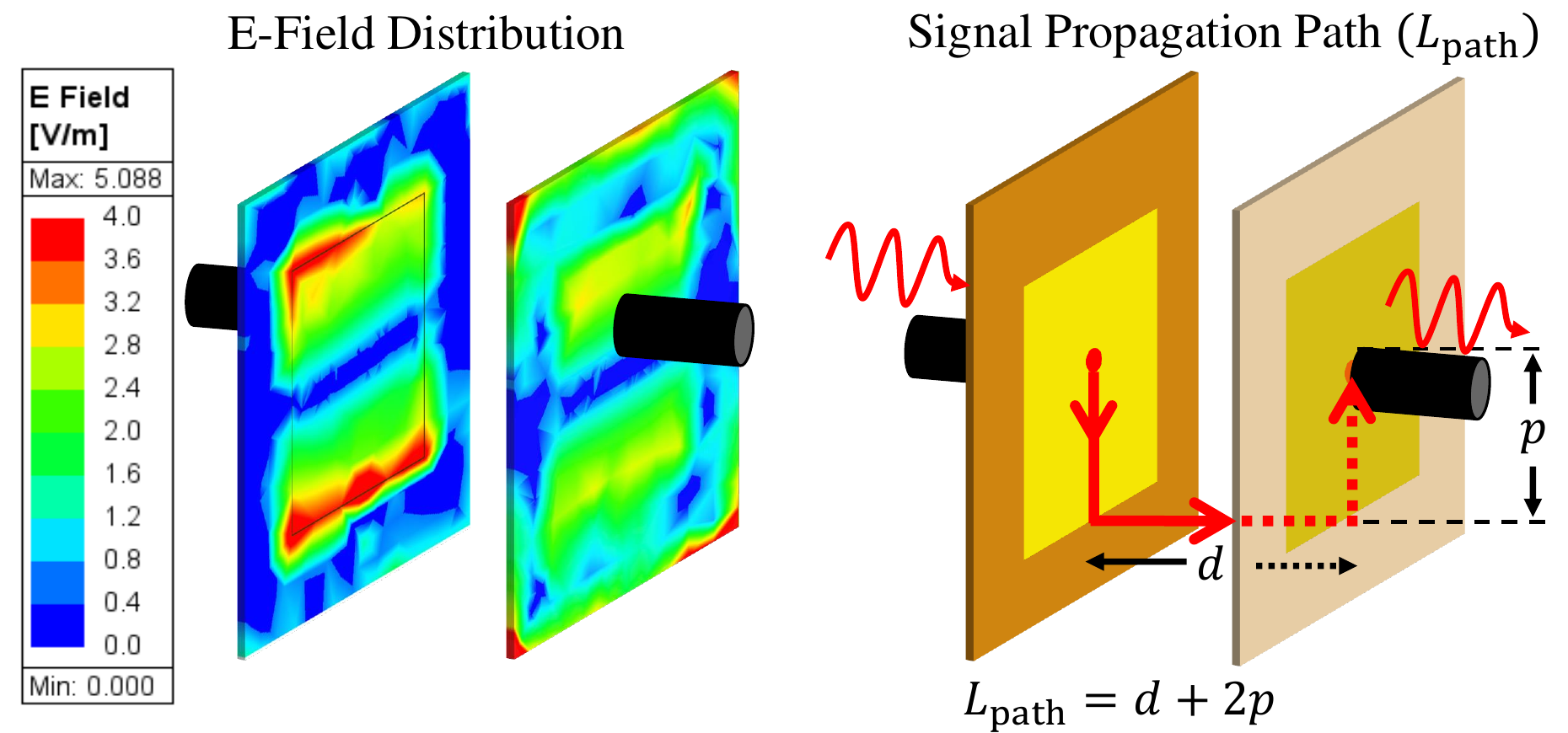}
\caption{Definition of signal path $L_{\rm{path}}$ relevant to insertion loss.}
\label{fig:insersion_loss_decrease}
    \end{center}
\end{figure}

To achieve a more compact form factor and enhance bandwidth performance to levels comparable with traditional RF connectors, the proposed design incorporates a narrow slot in the patch. This slot acts as a capacitive load, which, when introduced into a \(\lambda_e/4\) short-circuited transmission line, disrupts the continuity of current flow and shifts the resonant frequency to a lower range. Consequently, the effective electrical length of the short-circuited transmission line is reduced from \(\lambda_e/4\) to approximately \(\lambda_e/20\). This modification also suppresses the formation of standing waves, thereby mitigating resonance effects. Furthermore, the phase shift introduced by the capacitive load improves impedance matching over a broader frequency range, effectively extending the bandwidth from DC to RF.

\subsubsection{Low Loss Design}
\label{sec:Low Loss}
Fig.~\ref{fig:insersion_loss_decrease} (left) shows the electric field distribution for a traditional face-to-face square patch design. The electric field intensity is weakest at the center and reaches its maximum near the edges, indicating that radiation and coupling between the two patch antennas predominantly occur along the top and bottom radiating edges. From this electric field distribution, the total signal propagation path length, \(L_{\rm{path}}\), from the input port of the left antenna to the output port of the right antenna (related to \(S_{12}\)), can be decomposed into two components: the dielectric path length within each patch (denoted as \(p\)) and the free-space propagation path between the patches (denoted as \(d\)). Accordingly, the total signal propagation path from one antenna port to the other is given by \(L_{\rm{path}} = d + 2p\). Longer propagation distances \(L_{\rm{path}}\) inherently result in greater RF signal attenuation. Therefore, minimizing \(L_{\rm{path}} = d + 2p\) serves as an effective design guideline for RF connectors to achieve low insertion loss and maximize \(S_{12}\) performance.

As illustrated in Fig.~\ref{fig:connector_design_trend}, the initial design of the traditional square patch features a large air gap (\(d = 0.2\lambda_e\)), resulting in an extended propagation path \(L_{\rm{path}}\) and extremely poor \(S_{12}\) performance of \(-84.09\,\mathrm{dB}\). In stage 1, \(S_{12}\) improves significantly to \(-1.86\,\mathrm{dB}\) when the air gap \(d\) is reduced to zero, eliminating free-space propagation losses. Stage 2 further enhances \(S_{12}\) to \(-0.55\,\mathrm{dB}\), attributed to the reduction of dielectric path loss as some signals travel from the feed point to the patch corners. In stage 3, the introduction of a metallic via shortens the patch length from \(\lambda_e/2\) to \(\lambda_e/4\), effectively minimizing \(p\) and improving \(S_{12}\) to \(-0.47\,\mathrm{dB}\). In the final proposed design, the patch length is further reduced to \(\lambda_e/20\), achieving the smallest \(p\) and an \(S_{12}\) of \(-0.06\,\mathrm{dB}\). 

This progressive reduction of \(L_{\rm{path}}\) directly lowers insertion loss and substantially enhances \(S_{12}\) performance across the operating frequency band. It is worth noting that, for cost-effectiveness, the current design utilizes a lossy FR4 substrate with a relatively high loss tangent of 0.02. The \(S_{12}\) performance could be further improved by employing a low-loss substrate material.

\subsection{LEGO-Inspired Precise Alignment Design}
\label{sec:docking}
\begin{figure}[htbp!]
      \begin{center}
      \includegraphics[width=0.6\textwidth]
{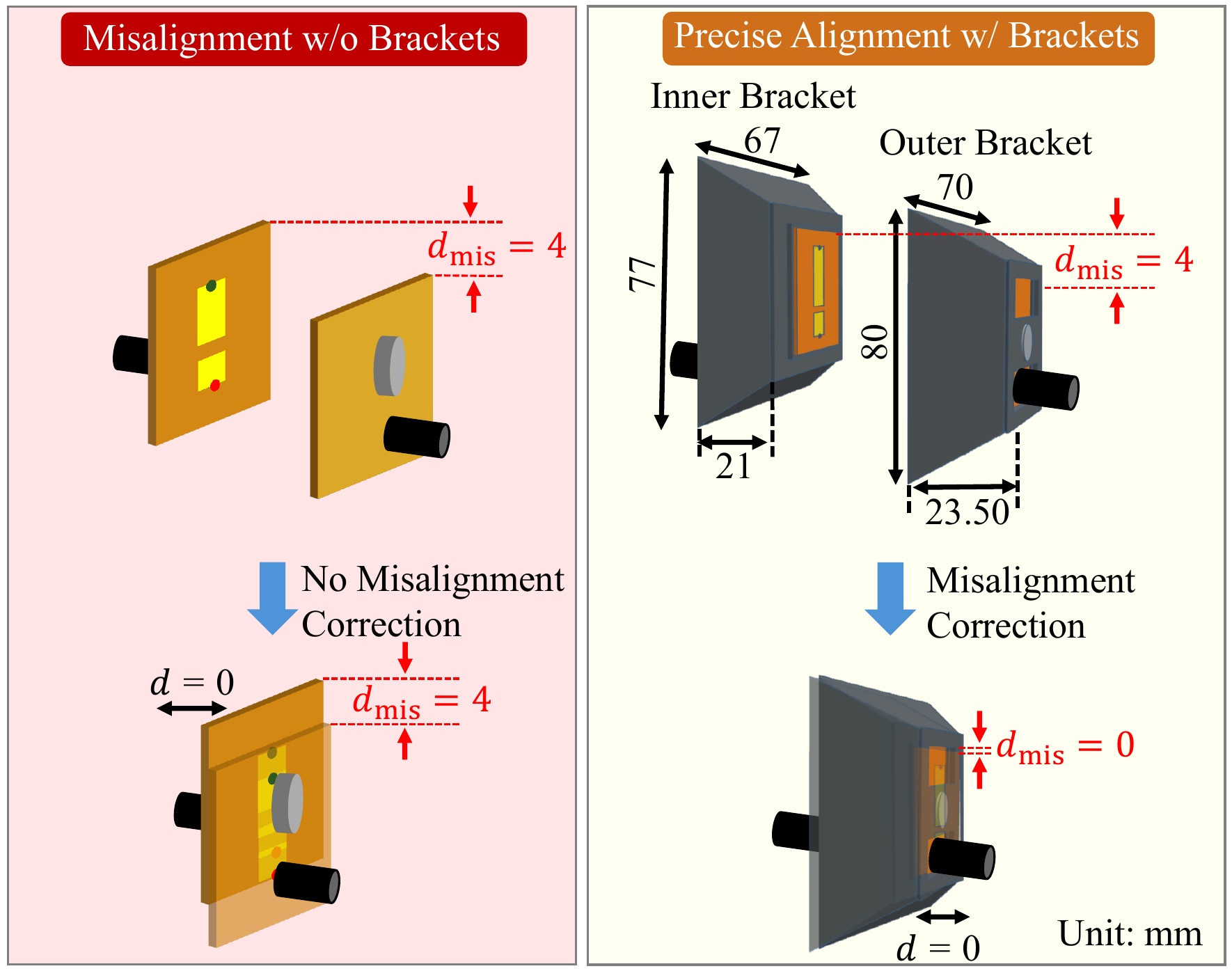}
\caption{Comparison of proposed RF connector alignment without (left) and with (right) the LEGO-inspired alignment brackets.}
\label{fig:connector_alignment}
    \end{center}
\end{figure}

\begin{figure}[htbp!]
      \begin{center}
      \includegraphics[scale=0.5]{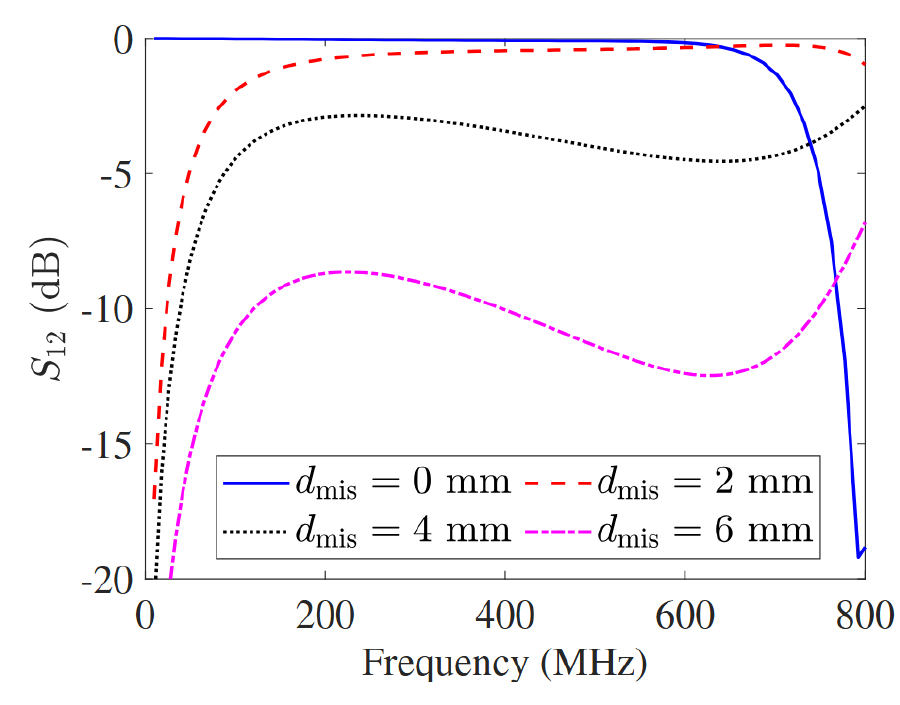}
      \caption{Impact of alignment on the ${S}_{12}$ of the proposed RF connector.}
\label{fig:S12_misalign}
      \end{center}
\end{figure}

Precise mechanical alignment is essential for maintaining optimal \(S_{12}\) performance in the proposed RF connector. As shown in Fig.~\ref{fig:connector_alignment} left, vertical misalignment between the two connector plates, denoted as \(d_{\text{mis}}\), can arise even when the air gap \(d\) is zero, if alignment correction is absent. The impact of such misalignment on \(S_{12}\) is quantitatively evaluated in Fig.~\ref{fig:S12_misalign} for varying \(d_{\text{mis}}\) levels. As \(d_{\text{mis}}\) increases from 0 to \(6\,\mathrm{mm}\), \(S_{12}\) deteriorates significantly across the operating frequency band. In the perfectly aligned case (\(d_{\text{mis}} = 0\,\mathrm{mm}\)), \(S_{12}\) remains close to \(0\,\mathrm{dB}\), indicating minimal insertion loss. In contrast, at \(d_{\text{mis}} = 6\,\mathrm{mm}\), \(S_{12}\) falls below \(-10\,\mathrm{dB}\), highlighting severe performance degradation caused by misalignment.

To address this issue, a pair of 3D-printed alignment brackets inspired by LEGO brick architecture is introduced, as shown in Fig.~\ref{fig:connector_alignment} right. The alignment system comprises a concave outer bracket and a convex inner bracket designed to mate in a guided manner. The RF connector board is mounted on the inward-facing surface of the concave bracket and the outward-facing surface of the convex bracket. This bracketed design, integrated with embedded magnets, ensures precise alignment of the connector plates along all axes, effectively reducing the misalignment distance \(d_{\text{mis}}\) to zero. The proposed RF connector utilizes a passive mechanical interlock that enables repeatable and tool-free alignment.

\section{Phased Array Antenna Design}
\label{sec:array_design}
Our previous work~\cite{10818494} demonstrated a reconfigurable UAV swarm-based Yagi–Uda antenna. While it exhibited beamforming capability, it lacked beam steering functionality due to the use of only a single driven element, with all other elements functioning as parasitic directors and reflectors. In this work, we propose a UAV swarm-based phased array system capable of achieving true beam steering functionality. This section presents the antenna array design and evaluates the effects of UAV structural components, internal circuitry, and the docking mechanism on the array’s radiation pattern through simulations. Additionally, we analyze the scalability of the proposed phased array in terms of achievable gain and operating frequency.

\subsection{Antenna Array Design}
\label{sec:antenna_design}
We adopt an antenna structure similar to the Yagi-Uda antenna, as it can be lightweight and has minimal resistance to wind, both from the UAV propellers and from the surroundings. Each element of the phased array is designed to be an antenna with one reflector and one driven element. The driven element is designed to have a length of $0.51\lambda$. The length of the reflector element is $0.53\lambda$, which is approximately 5\% longer than the driven element to effectively reflect electromagnetic energy toward the directors. The distance between the reflector element and the driven element is $0.29\lambda$ and the array element spacing is $0.5\lambda$. Beam steering in the phased array is achieved by exciting each driven element with a phase-shifted signal. In a linear array of $N$ elements, the phase of each element can be calculated using \eqref{eq:phaseshift}. 
\begin{equation}
   \Phi_{n} = (n-1)kd_{\rm{ele}}\cos\theta_{\rm{steer}},\quad n = 0,1,2,\dots,N
    \label{eq:phaseshift}
\end{equation}
where $k$ is the wavenumber defined as $2\pi/\lambda$, $\lambda$ is the wavelength, $d_{\rm{ele}}$ is the distance between two adjacent array elements, as depicted in Fig.~\ref{fig:blockage} (left), and $\theta_{\rm{steer}}$ is the direction of the main beam, as shown in Fig.~\ref{fig:Gain_diff_UAVs} (left).

\subsection{Blockage Effect on Radiation Pattern}
\label{sec:sim_blockage}
\begin{figure}[htbp!]
      \begin{center}
      \includegraphics[width=0.4\textwidth]
{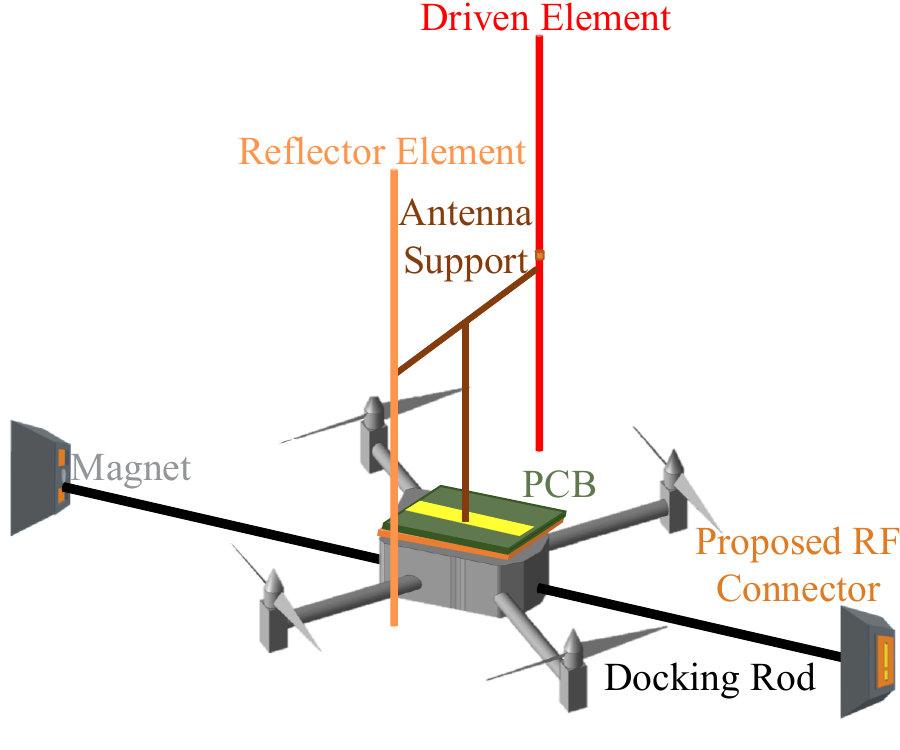}
\caption{Simulation model of a UAV swarm-based phased array element with the proposed RF connectors.}
\label{fig:model}
    \end{center}
\end{figure}

To assess the influence of the UAV platforms and the proposed RF connectors on the antenna radiation characteristics, we model the UAV-mounted array element with the proposed RF connector docking mechanism in HFSS as in Fig.~\ref{fig:model}.

The driven and reflector elements of the antenna are modeled as copper rods. The antenna support structure is modeled using polylactic acid (PLA), while the body and propellers of the UAV are modeled using polyester. The internal circuitry of the UAV, along with the signal generator mounted on top, is represented as a printed circuit board (PCB) consisting of copper microstrip lines on an FR4 epoxy substrate above a copper ground plane. Neodymium magnets are used to implement the magnetic docking system described in Section~\ref{sec:docking}, due to their strong magnetic properties. The RF connector patch, modeled as a silver strip with a resistivity of \( 1.27 \times 10^{-4}~\Omega\cdot\mathrm{m} \)~\cite{voltera_vone}, is placed on an FR4 substrate with a copper ground. The alignment brackets of the proposed RF connectors are modeled using polyester. The docking rods, which connect UAVs and maintain a fixed inter-element spacing $d_{\rm{ele}}$, are modeled using carbon fiber with a conductivity of $1\times10^{4}$~S/m, as reported in~\cite{khan2019experimental}.

\begin{figure}[ht]
    \centering
    \includegraphics[width=0.5\textwidth]{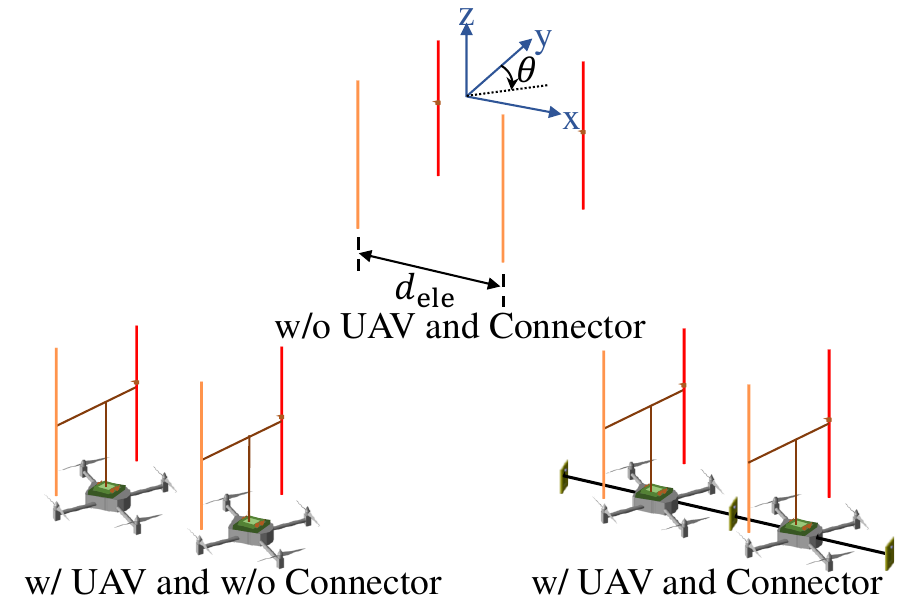}
    \hfill
    \includegraphics[width=0.4\textwidth]{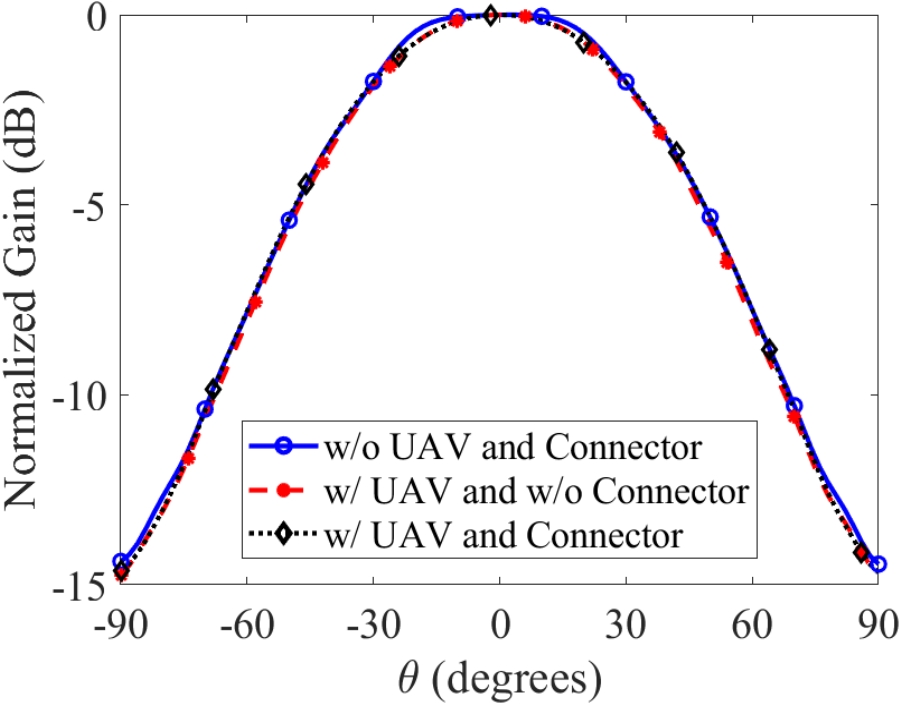}
    \caption{ Structural configurations of a two-element phased array to evaluate the blockage effects of the UAV body and the proposed RF connector on the antenna radiation pattern (left). Simulated H-plane radiation patterns for the three configurations in (a) (right).}
    \label{fig:blockage}
\end{figure}

We consider three configurations of a two-element phased array as illustrated in Fig.~\ref{fig:blockage} (left): (i) only the array antenna without the UAV and the RF connectors, (ii) the antenna mounted on the UAVs, but without the docking rods and the RF connectors, and (iii) the antenna mounted on the UAVs with the docking rods and the RF connectors. Fig.~\ref{fig:blockage} (right) presents the corresponding simulated radiation patterns in the H-plane for each case. The comparison highlights how the UAV body and magnetic structures affect overall radiation performance. The main beam direction, gain and beamwidth remain largely preserved, confirming that the blockage effects of UAV body, antenna support and RF connectors have minimal effect on the radiation performance of the antenna array.

\subsection{Reconfigurable Gain}
\begin{figure}[ht!]
    \centering
    \includegraphics[width=0.4\textwidth]{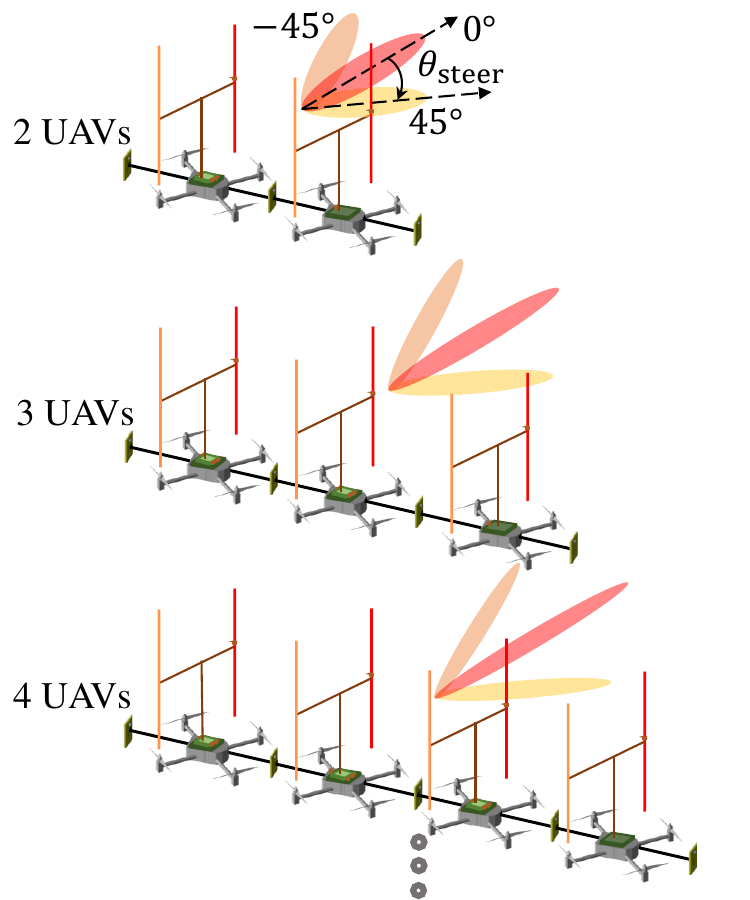}
    \hfill
    \includegraphics[width=0.4\textwidth]{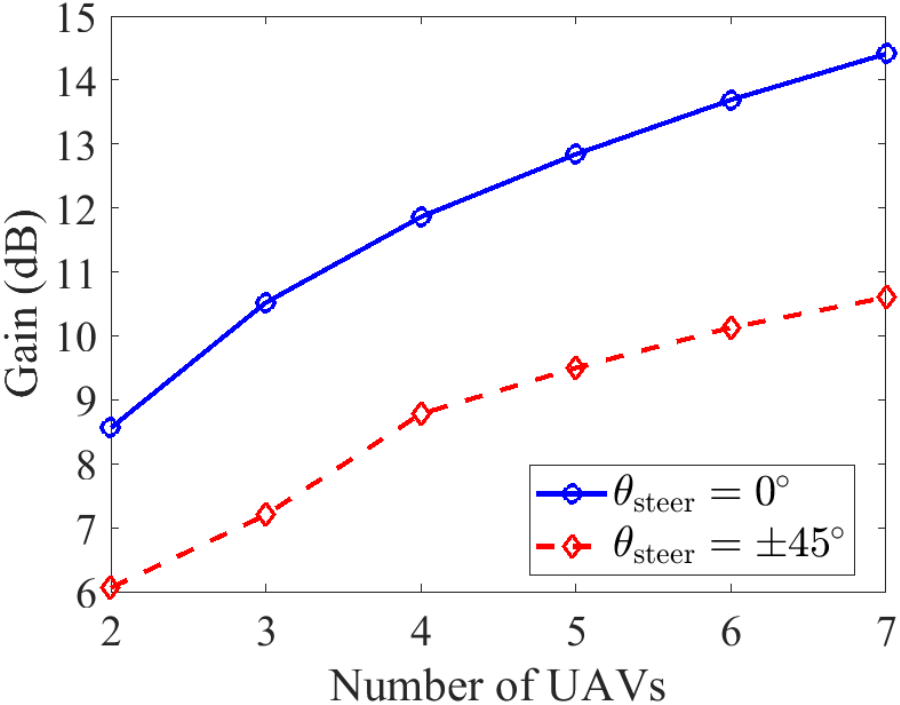}
    \caption{ UAV swarm-based array with increasing UAV count (left). Variation of antenna gain with UAV count (right).}
    \label{fig:Gain_diff_UAVs}
\end{figure}

The UAV swarm-based array provides flexibility and adaptability through its dynamic architecture. In the proposed system, each UAV module carries a single antenna module as illustrated in Fig.~\ref{fig:Gain_diff_UAVs} (left), forming part of a larger phased array structure.

The integration of additional UAV modules via the proposed RF connectors dynamically reconfigures the performance parameters, such as array gain and beamwidth. As depicted in Fig.~\ref{fig:Gain_diff_UAVs} (right), the array gain increases from approximately $8.7$ dB with two UAVs to $14.41$ dB with seven UAVs  for $\theta_{\rm{steer}}$ at $0^\circ$ and from $6.01$ dB to $10.54$ dB for $\theta_{\rm{steer}}$ at $\pm 45^\circ$. This modular approach enables scalable deployment, allowing dynamic control of array gain by adjusting the number of UAVs in formation while maintaining consistent beam steering capability. 

%The modular nature of the system reduces the logistical burden associated with transporting and assembling large antenna arrays. Each UAV can be independently deployed and magnetically docked mid-flight. This functionality facilitates rapid assembly of UAV swarm-based phased arrays, an essential feature for time-sensitive operations and dynamic wireless communication environments.

\subsection{Different Operating Frequency Design}
\begin{figure}[htbp]
\centering
\includegraphics[width=0.65\textwidth]{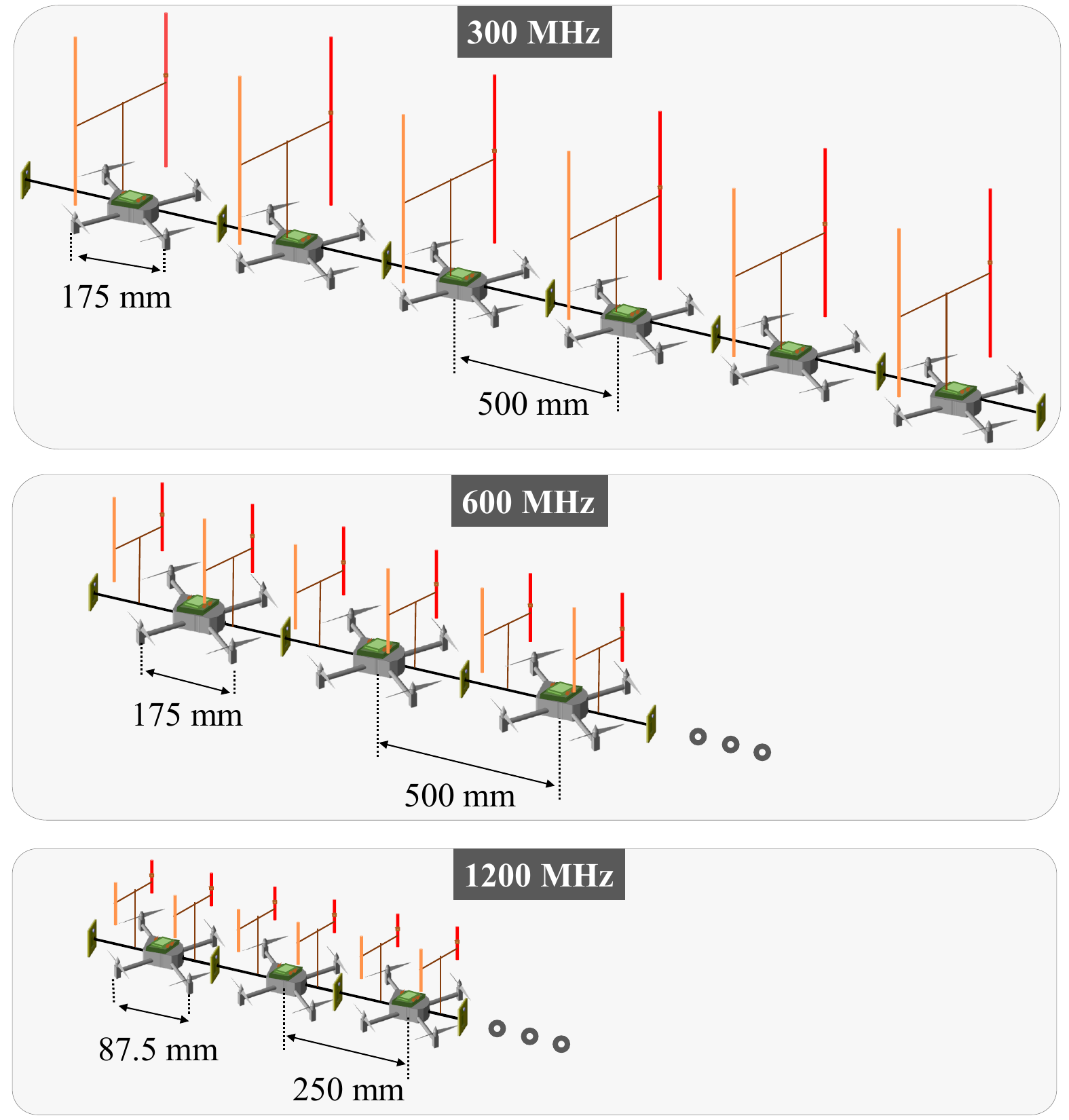}
\caption{Reconfigurability of UAV swarm-based array across frequency.}
\label{fig:freq}
\end{figure}

This section evaluates the scalability of the proposed UAV swarm-based phased array with the RF connector across different operating frequencies. As depicted in Fig.~\ref{fig:freq}, the number of array elements per UAV and their placement are determined by the element size and the inter-element spacing, \(d_{\rm{ele}}\). A trade-off exists between the desired \(d_{\rm{ele}}\) and the minimum safe UAV-to-UAV spacing needed to mitigate aerodynamic interference within the swarm. Previous studies indicate that maintaining a UAV-to-UAV spacing greater than twice the propeller diameter effectively reduces aerodynamic turbulence from neighboring UAVs and ensures stable swarm flight~\cite{OO2023108535}. For instance, a UAV with a length of 175\,mm and a propeller diameter of 127\,mm can maintain a spacing of \(d_{\rm{ele}} = 500\,\mathrm{mm}\), which corresponds to half the wavelength at 300\,MHz. At 600\,MHz, the same UAV can support two array elements while preserving this spacing. For operation at 1200\,MHz, smaller UAVs with a length of 87.5\,mm, each carrying two elements and maintaining \(d_{\rm{ele}} = 250\,\mathrm{mm}\), satisfy the aerodynamic constraints. These findings demonstrate that increasing the array element density per UAV and reducing UAV dimensions are effective strategies for scaling the design to higher operating frequencies.

\begin{figure}[h!]
\centering
\includegraphics[width=0.5\textwidth]{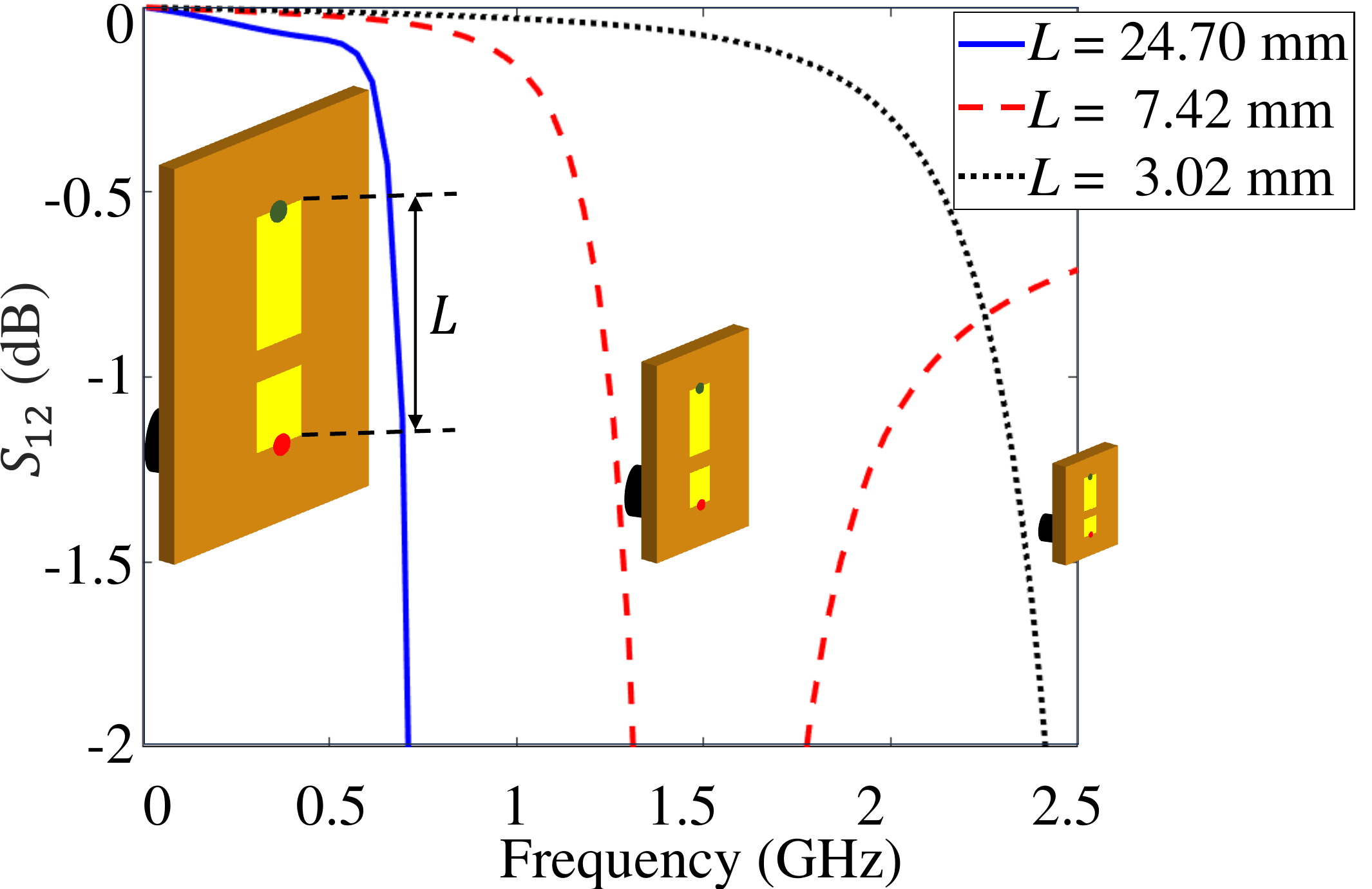}
  %\caption{Reconfigurability of the magnetic RF connector over frequency}
\caption{Reconfigurability of the proposed RF connector across frequency.}
\label{fig:freq_magrf}
\end{figure}

Fig.~\ref{fig:freq_magrf} shows that decreasing the patch length $L$ of the RF connector increases the maximum frequency for which the RF connector can be used effectively. Considering the low insertion loss of $0.1$ dB, an RF connector with $L = 24.70$ mm can be used up to $0.7$ GHz frequency. $L$ can be scaled down to $7.42$ mm to reach a maximum operating frequency of approximately $1.4$ GHz. For a maximum operating frequency of $2$ GHz, $L$ can be scaled down to $3.02$ mm.

The frequency-based analysis presented in Fig.~\ref{fig:freq} and Fig.~\ref{fig:freq_magrf} confirms that the proposed method is scalable across a wide range of operating frequencies. This scalability enables adaptability for diverse applications, ranging from low-frequency operations, such as long-range communication and ground-penetrating systems, to high-frequency applications, including satellite communications, high-resolution imaging, and precision sensing.

\begin{figure}[!t]
\centering
{\includegraphics[width=0.42\textwidth]{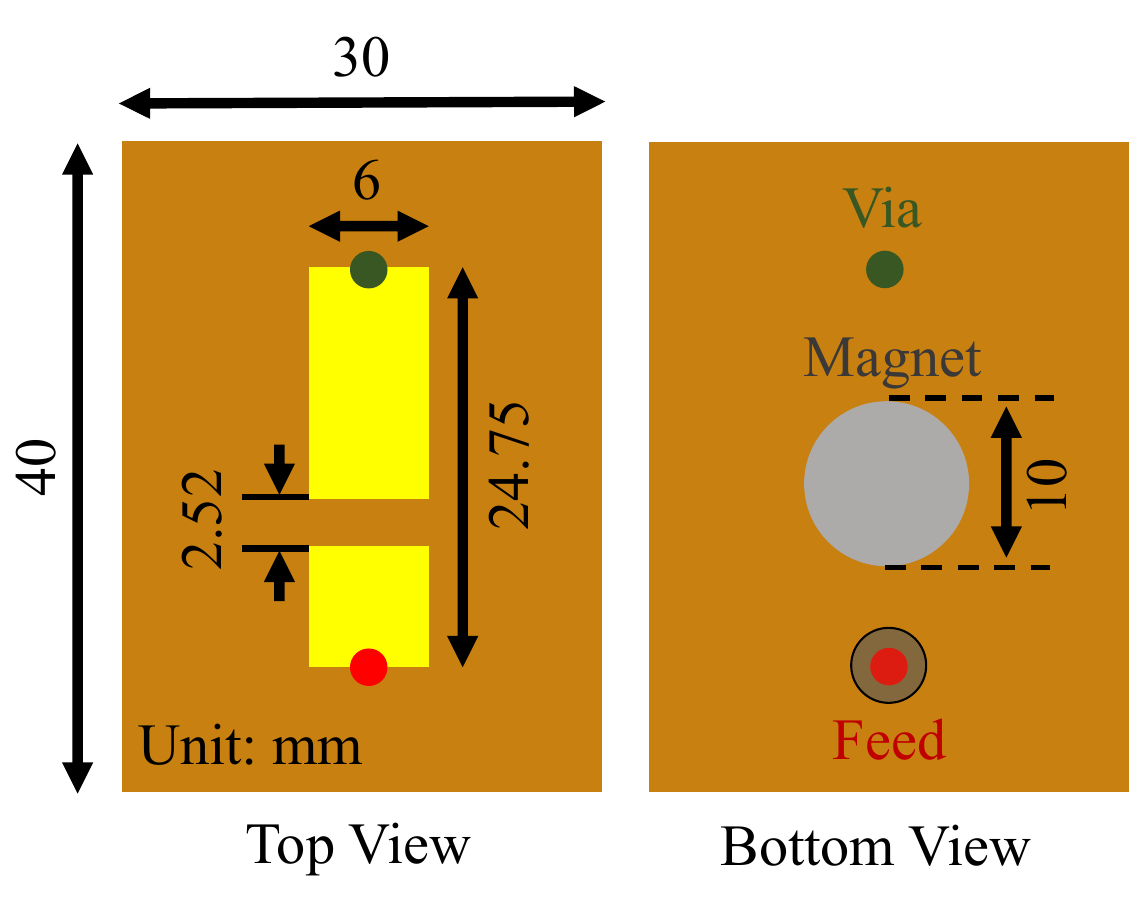}}
\caption{ Geometry and dimensions of the proposed RF connector.}
\label{fig:rf_connector_dim}
\end{figure}

\begin{figure}[htbp]
\centering
\includegraphics[width=0.6\textwidth]{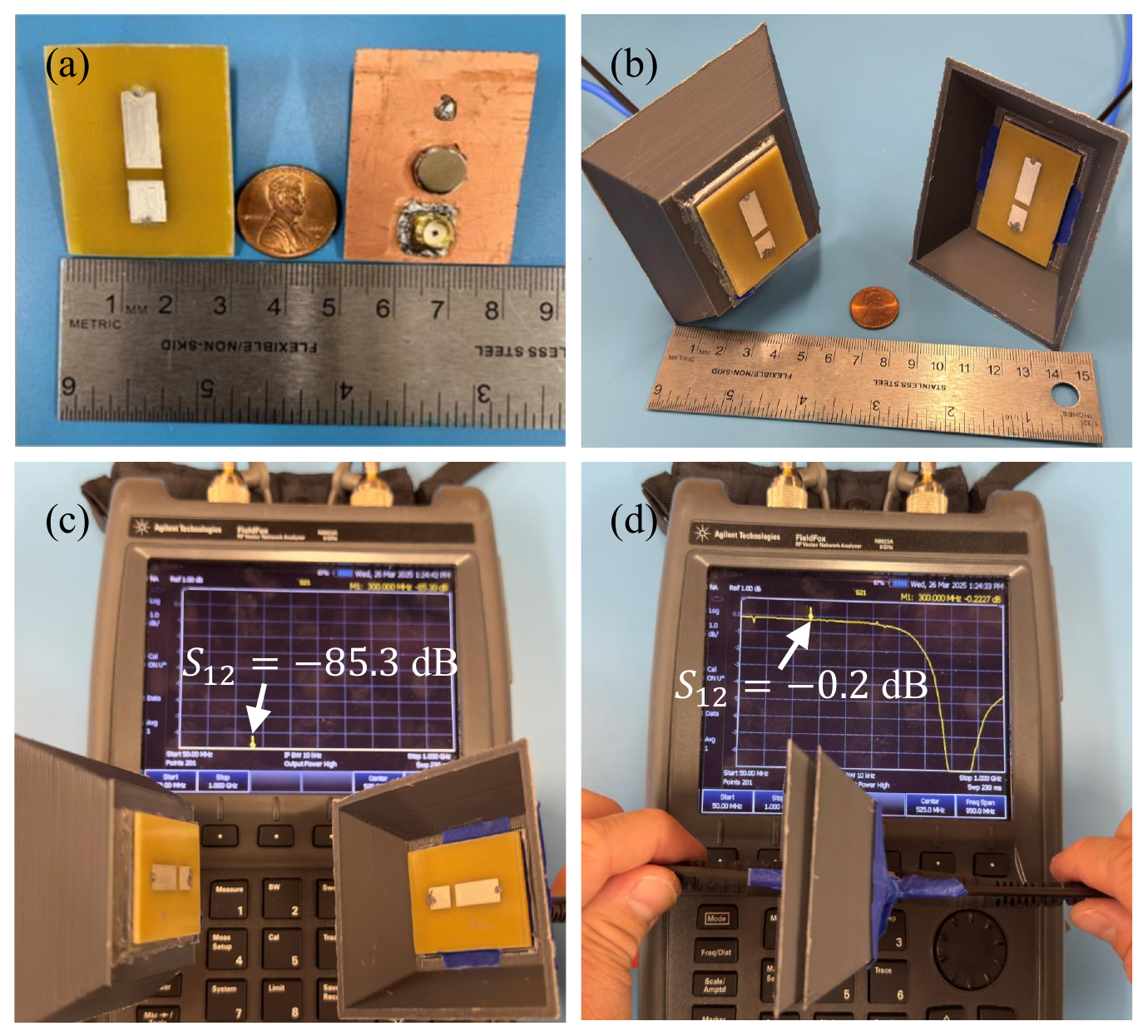}
%\caption{Fabricated magnetic RF connector with alignment brackets and measurement setup for testing signal path loss.}
\caption{Fabrication of the proposed RF connector: (a) Fabricated connector components. (b) Connectors with LEGO-inspired alignment brackets. (c) $S_{12}$ measurement of the connectors before, and (d) after connection establishment. The video of transitioning from (c) to (d) is provided in the supplemental documentation.}
\label{fig:magnetic_connector_grid}
\end{figure}

\begin{figure}[htb]
\centering
\includegraphics[width=0.48\textwidth]{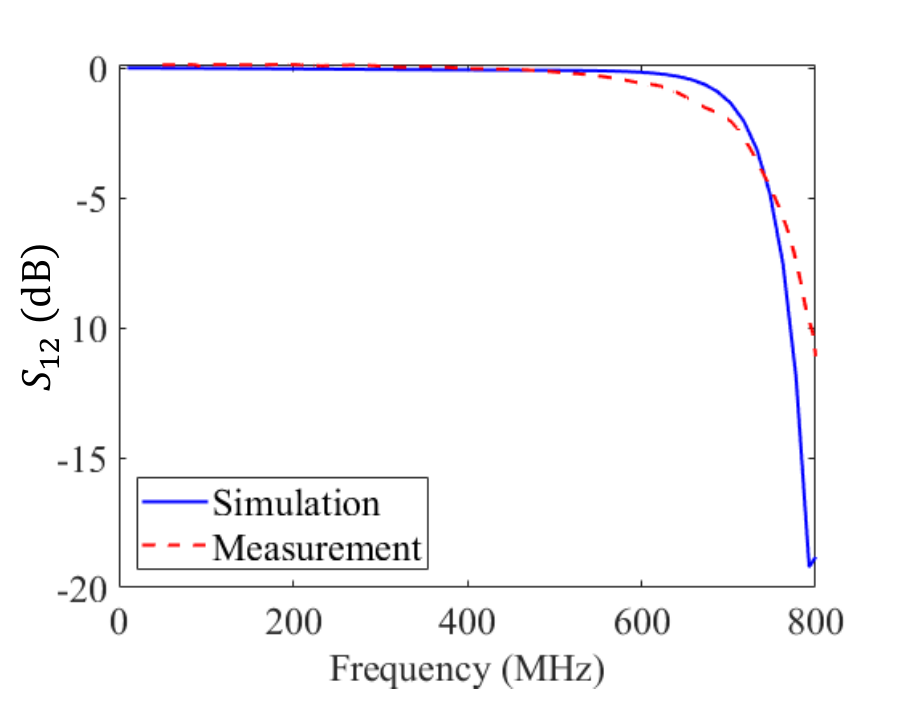}
\caption{ Measured $S_{12}$ of the proposed RF connectors after establishing face-to-face contact.}
\label{fig:mag_RF_S12}
\end{figure}

\section{Experimental Demonstration}

This section presents the experimental validation of the proposed RF connector and UAV swarm-based phased array, bridging theoretical design and practical implementation. Measured S-parameters and radiation patterns from a non-flying prototype tested in an anechoic chamber are compared with simulation results to verify the system’s performance. Furthermore, the docking and undocking capabilities of two flying UAVs, as well as beam steering functionality using the flying prototype, are experimentally demonstrated.

\subsection{Proposed RF Connector Fabrication and Measurement}
The RF connector is manufactured according to the dimensions given in Fig.\ref{fig:rf_connector_dim}. It consists of a rectangular slotted patch of silver \cite{voltera_vone} printed on an FR4 substrate. There is a copper sheet underneath the substrate that acts as a ground plane. An SMA connector is soldered at one end of the patch, and a metal via on the other end of the patch to short the patch to ground. A $2$ mm-thick neodymium magnet is attached to the back of each connector. Fig.\ref{fig:magnetic_connector_grid}(a) shows the fabricated RF connectors. The alignment brackets in Fig. \ref{fig:magnetic_connector_grid}(b) ensure precise alignment between the two RF connector patches. Subfigures (c) and (d) in Fig.~\ref{fig:magnetic_connector_grid} show the $S_{12}$ performance of the RF connectors before and after establishing the connection. The $S_{12}$ at the operating frequency of $300$ MHz is $-0.2$ dB, consistent with the simulation result in Fig.~\ref{fig:mag_RF_S12}. This indicates that the RF signal experiences negligible insertion loss when propagating through the proposed connectors. No interference was observed in the $S_{12}$ performance from the magnets, confirming the effectiveness of magnet type and placement for stable UAV operation and reliable RF connectivity \cite{zheng2021unmanned}.

\subsection{Non-Flying UAVs Chamber Measurement}

\begin{figure}[h!]
\centering
{\includegraphics[width=0.5\textwidth]{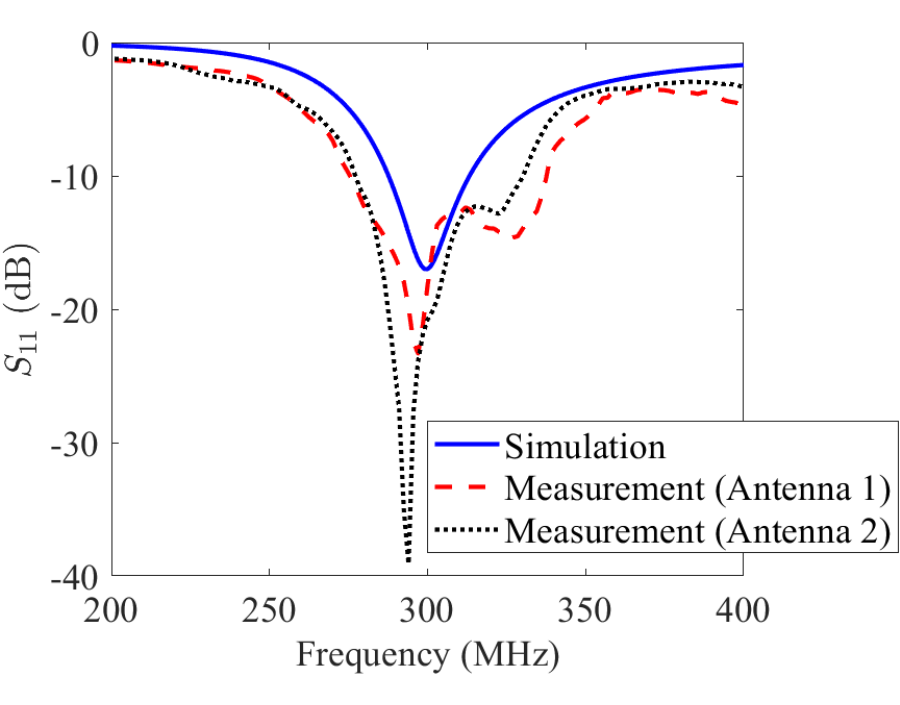}}
\caption{Measured $S_{11}$ of the fabricated antennas of the UAV swarm-based two-element phased array.}
\label{fig:S11_yagi}
\end{figure}

To evaluate the beam steering performance of the proposed UAV swarm-based phased array illustrated in Fig.~\ref{fig:beamsteering_patterns} (left), a measurement setup was established inside an anechoic chamber, as shown in Fig.~\ref{fig:beamsteering_patterns} (right). The two UAVs were mounted on a low-scattering polystyrene block to minimize reflections and blockage effects during testing. The element spacing was set to \(d_{\rm{ele}} = 500\,\mathrm{mm}\) to achieve low sidelobe levels. Each UAV carried one element of the phased array, comprising a driven element with a length of \(510\,\mathrm{mm}\) and a reflector element with a length of \(530\,\mathrm{mm}\) long, both fabricated from copper wire with a diameter of \(2\,\mathrm{mm}\). A single RF signal generator was used to feed both antennas via an RF splitter: one output was connected to antenna~1 through phase shifter~1, while the other output was routed through the RF connectors and phase shifter~2 to antenna~2. The measured \(S_{11}\) performance of the fabricated antennas, presented in Fig.~\ref{fig:S11_yagi}, shows good agreement with the simulated results derived from the antenna model discussed in Section~\ref{sec:sim_blockage}.

\begin{figure}[ht!]
    \centering
    \includegraphics[width=0.48\textwidth]{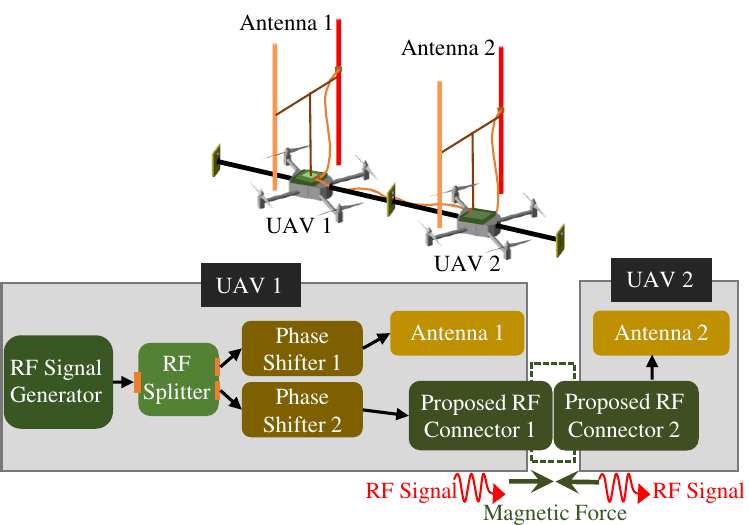}
    \hfill
    \includegraphics[width=0.48\textwidth]{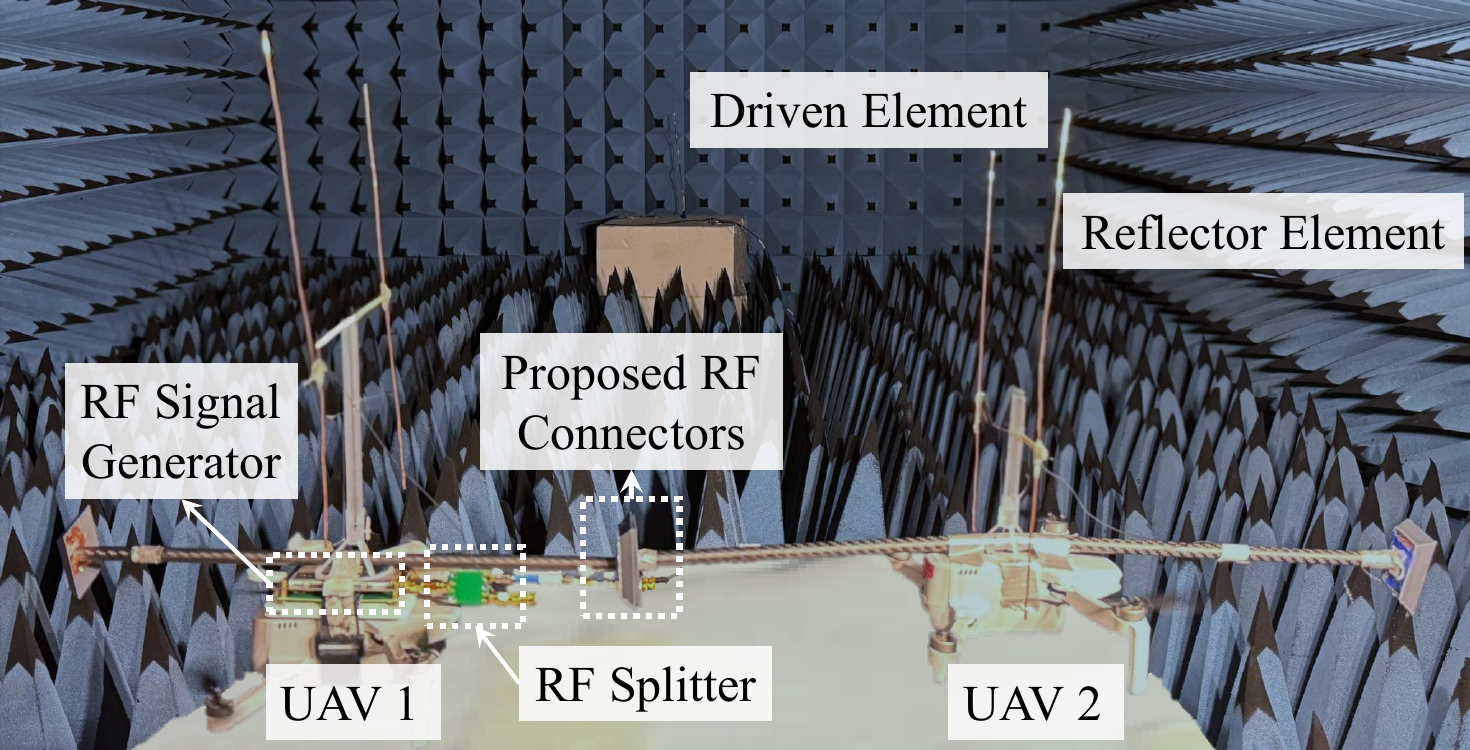}
    \caption{ Schematic of a two-element phased array (left). Beam steering measurement in the anechoic chamber (right).}
    \label{fig:beamsteering_patterns}
\end{figure}

\begin{figure}[htb!]
\centering
%\subfigure[$\theta_{\mathrm{steer}} = -45^\circ$]
\includegraphics[width=0.4\textwidth]{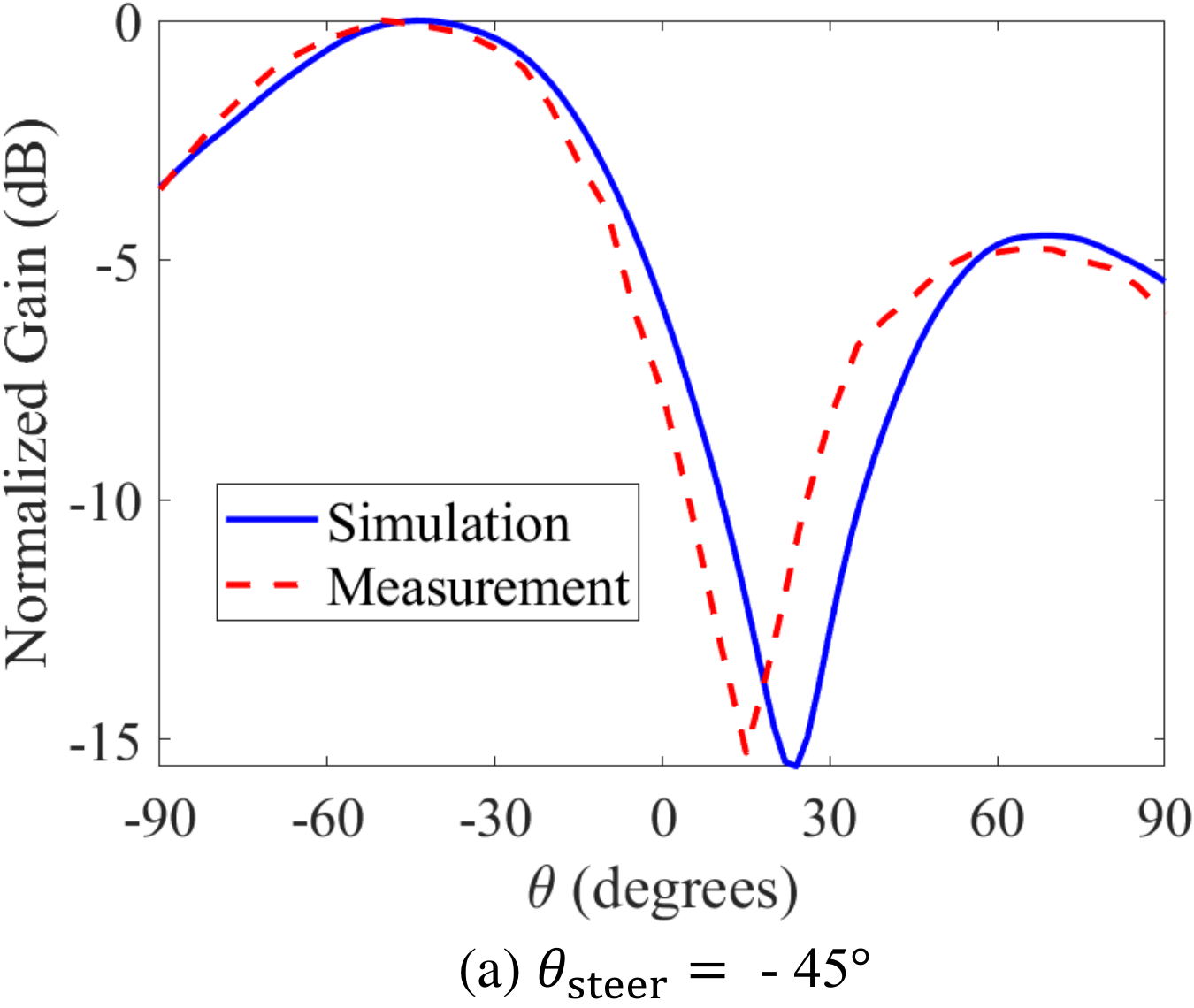}\label{fig:beamsteering_m45}
\hfill
%\subfigure[$\theta_{\mathrm{steer}} = 0^\circ$]
\includegraphics[width=0.44\textwidth]{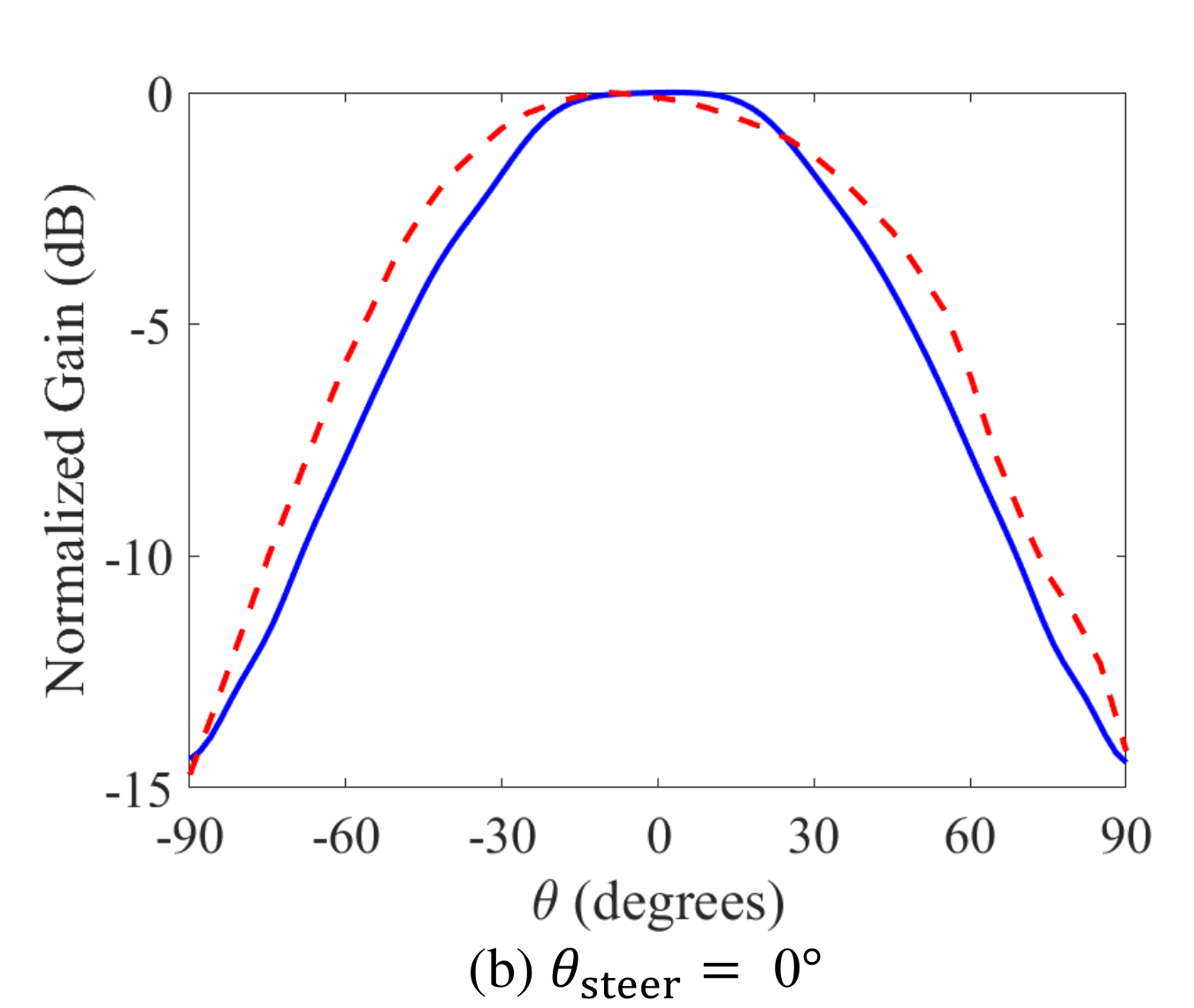}\label{fig:beamsteering_boreside}
\hfill
%\subfigure[$\theta_{\mathrm{steer}} = 45^\circ$]
\includegraphics[width=0.44\textwidth]{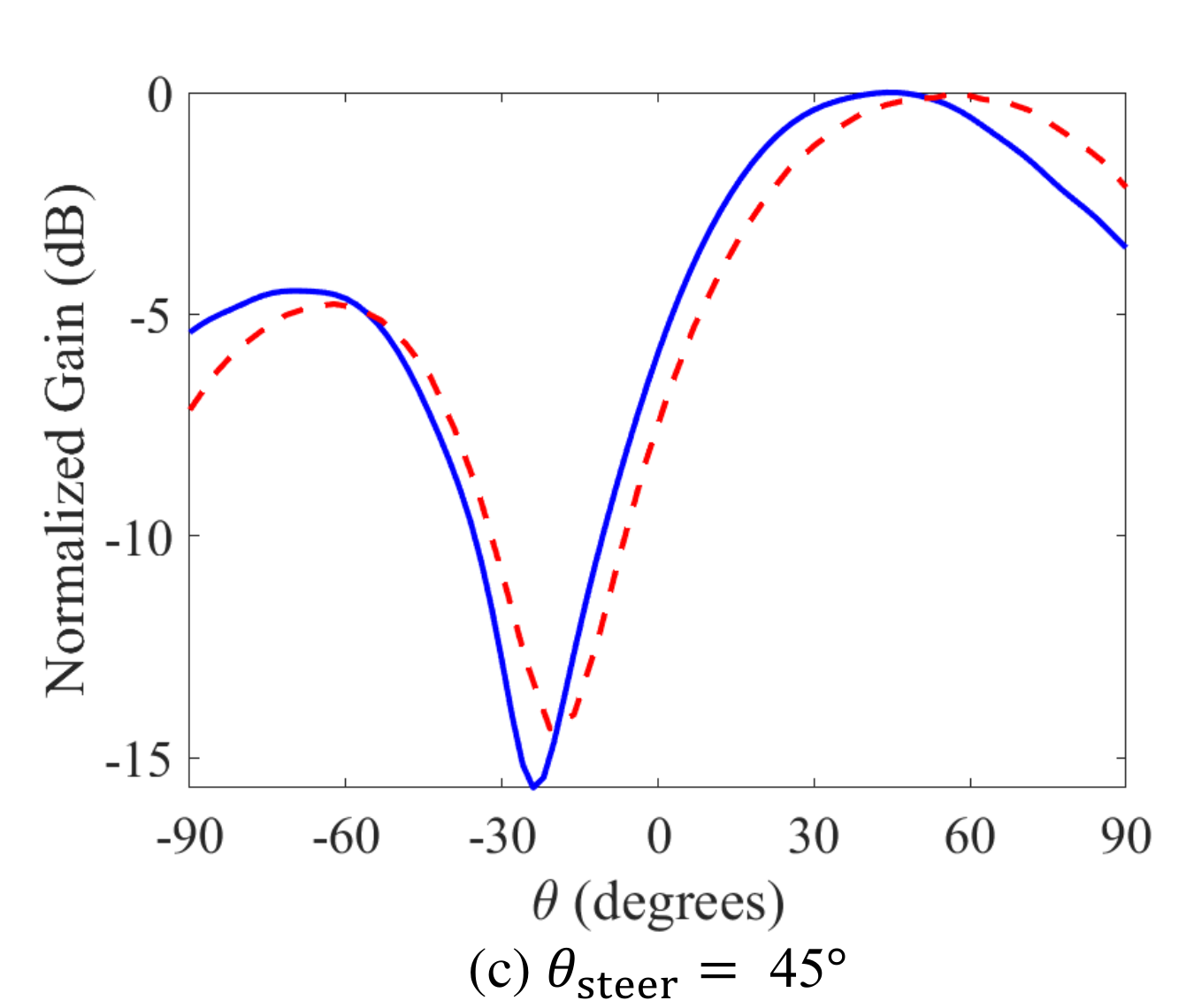}\label{fig:beamsteering_45}
\caption{Measured H-plane radiation patterns for various $\theta_{\rm{steer}}$.}
\label{fig:beamsteering_patterns}
\end{figure}

Fig.~\ref{fig:beamsteering_patterns} shows the measured radiation patterns for the $\theta_{\rm{steer}}$ directions \(-45^\circ\), \(0^\circ\), and \(45^\circ\), achieved by adjusting the phase $\Phi_{2}$ of the signal feed to antenna~2 following~\eqref{eq:phaseshift}. When $\Phi_{2}$ is zero, the array generates a main beam directed at \(0^\circ\), as shown in Fig.~\ref{fig:beamsteering_patterns}(b). By introducing a controlled $\Phi_{2}$, the main beam is steered to \(-45^\circ\) and \(45^\circ\), as illustrated in Figs.~\ref{fig:beamsteering_patterns}(a) and (c), respectively. The measurement results closely resemble the simulation results, which validate the effectiveness of the proposed RF connectors docking mechanism in enabling a continuous RF path and reliable beam steering in a UAV swarm-based phased array. 

\begin{figure}[h!]
\centering
\includegraphics[width=0.55\textwidth]{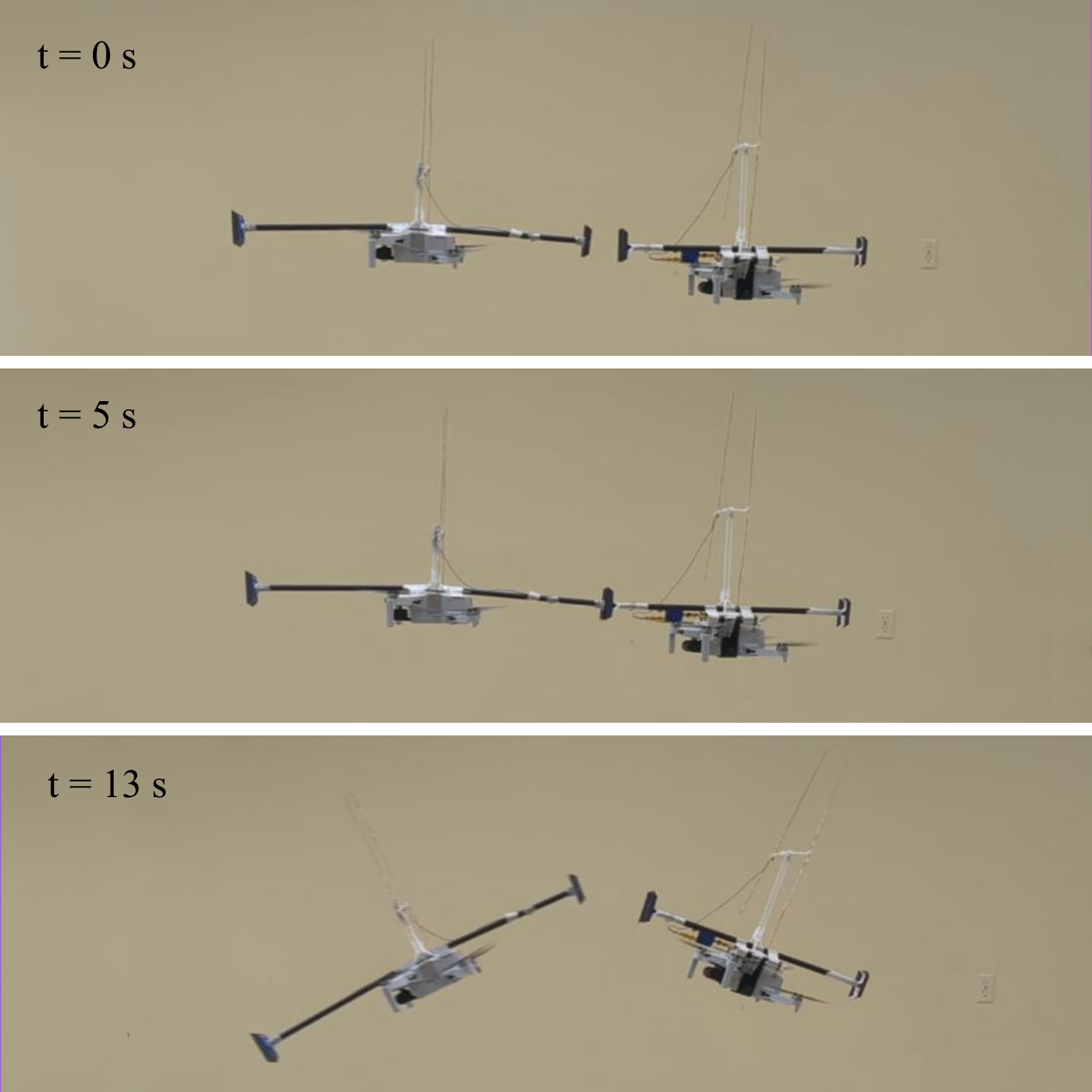}
\caption{A demonstration of mid-flight UAV docking using the proposed RF connector was successfully completed at 5 seconds, with undocking achieved at 13 seconds by commanding the UAVs to separate. The video is provided in the supplemental materials.}
\label{fig:docking_demo}
\end{figure}

\subsection{Flying UAVs Measurement}
The mid-flight docking of two UAVs is demonstrated in Fig.~\ref{fig:docking_demo}. When \(t = 0\,\text{s}\),
the UAVs approach each other, initiating docking action. At \(t = 5\,\text{s}\), the UAVs successfully dock via the proposed RF connector mounted at the ends of the carbon fiber docking rods, establishing a physical RF path. At \(t = 13\,\text{s}\), the UAVs are separated by flying in opposite directions to each other, thereby successfully disengaging the RF connectors. This docking and undocking demonstration validates the feasibility of a flying, reconfigurable UAV swarm-based phased array antenna.

\begin{figure}[h!]
\centering
{\includegraphics[width=0.55\columnwidth]{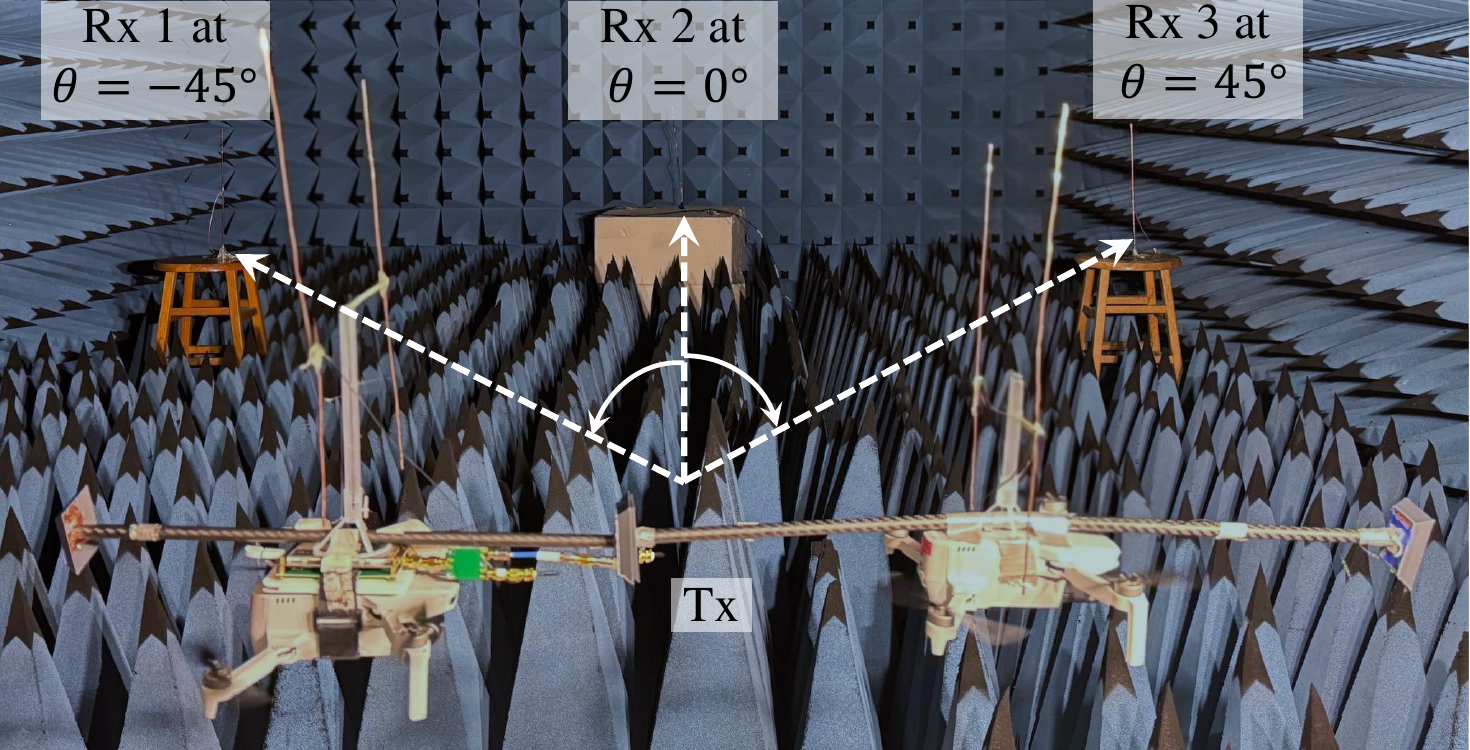}}
\caption{Beam steering measurement setup with flying UAVs.}
\label{fig:flying_mes_setup}
\end{figure}

\begin{figure}[h!]
\centering
\includegraphics[width=0.5\textwidth]{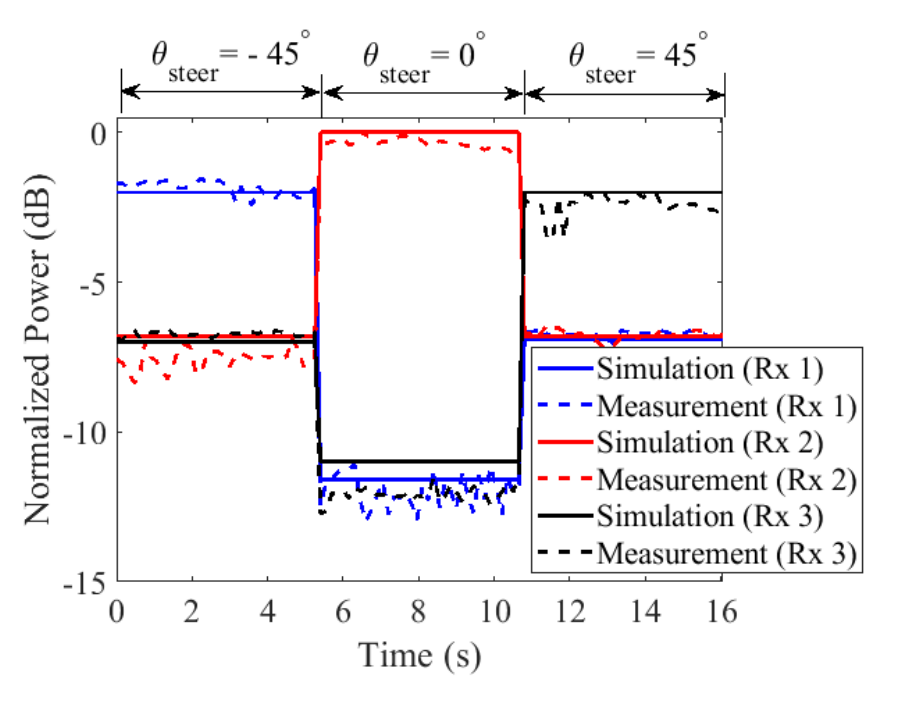}
\caption{Measured received power with the receivers in Fig.~\ref{fig:flying_mes_setup} at different $\theta_{\rm{steer}}$ directions.}
\label{fig:Flying_beamsteering_patterns}
\end{figure}

The beam steering capability of the flying UAV swarm-based phased array is validated using the experimental setup illustrated in Fig.~\ref{fig:flying_mes_setup}. Three receivers, Rx~1, Rx~2, and Rx~3, equal distant from the center of the array, are placed at \(\theta = -45^\circ\), \(0^\circ\), and \(45^\circ\), respectively. The phase of the feed signal to antenna~2, \(\Phi_2\), is adjusted to sequentially steer the main beam toward each receiver.

Fig.~\ref{fig:Flying_beamsteering_patterns} presents the normalized received power at each receiver as the beam is steered to \(\theta_{\rm{steer}} = -45^\circ\), \(0^\circ\), and \(45^\circ\). When steered to \(-45^\circ\), Rx~1 receives the highest power with a fluctuation of only \(0.8\,\mathrm{dB}\), while Rx~2 and Rx~3 measure signals approximately \(5\,\mathrm{dB}\) lower, confirming the correct \(\theta_{\rm{steer}}\) direction. Similarly, when the beam is steered to \(0^\circ\), Rx~2 records the highest received power with a minimal variation of \(0.2\,\mathrm{dB}\). Finally, steering to \(45^\circ\) results in Rx~3 receiving the strongest signal, with power variation within \(1.5\,\mathrm{dB}\) over time. The low fluctuations observed in the measured received power across the three \(\theta_{\rm{steer}}\) directions validate that the proposed RF connector maintains a rigid mechanical connection between flying UAVs with negligible positional error.

\section{Conclusion}
In this work, we present a theory-to-practice solution for a UAV swarm-based phased array system and experimentally validate its beam steering capability. The core innovation lies in the MagSafe- and LEGO-inspired RF connector, enabling hands-free, precise structural docking and low-loss signal continuity without requiring inter-element phase synchronization. A multi-stage optimization of the RF connector achieves a compact form factor, DC-to-RF bandwidth, and a measured insertion loss as low as \(0.2\,\mathrm{dB}\). The scalability of the system is analyzed for varying array gains and operating frequencies. Experimental tests of the flying 2-element UAV swarm-based phased array system under anechoic chamber conditions show excellent agreement with simulation results, demonstrating reliable beam steering to \(-45^\circ\), \(0^\circ\), and \(45^\circ\) with minimal temporal fluctuations. The proposed UAV swarm-based phased array provides a practical, flexible, and rapidly deployable platform for advanced airborne communications, radar, and sensing applications.

%\bibliographystyle{unsrt}  
%\bibliography{main}  %%% Remove comment to use the external .bib file (using bibtex).


\end{document}